\documentclass[11pt]{article}

\usepackage[hidelinks]{hyperref}

\usepackage{geometry}
\usepackage{graphicx}

\usepackage{amsmath,amssymb}
\usepackage{booktabs}
\usepackage{multirow}
\usepackage{pdflscape} 
\usepackage{afterpage} 
\usepackage{makecell}  
\usepackage{setspace}  
\usepackage{array}

\usepackage{bm}
\usepackage{bbm}
\usepackage[detect-all]{siunitx}
\usepackage{subcaption}
\usepackage[inline]{enumitem}
\usepackage{makecell}
\usepackage{IEEEtrantools}
\usepackage{etoolbox}
\makeatletter
\patchcmd{\@IEEEeqnarray}{\relax}{\relax\intertext@}{}{}
\makeatother

\DeclareMathOperator{\kron}{\,\otimes\,}
\usepackage{xspace}
\newcommand*{\eg}{e.g.\@\xspace}
\newcommand*{\ie}{i.e.\@\xspace}

\newcommand{\sym}[1]{#1}

\newcolumntype{P}[1]{>{\raggedright\arraybackslash}p{#1}}

\usepackage[round]{natbib}
\usepackage{authblk}

\title{Improving Crime Count Forecasts Using Twitter and Taxi Data}
\author[1,*]{Lara Vomfell}
\author[2,3]{Wolfgang Karl~H\"ardle}
\author[2]{Stefan Lessmann}
\affil[1]{Warwick Business School, University of Warwick, Coventry, UK}
\affil[2]{Faculty of Business and Economics, Humboldt University of Berlin, Unter den Linden 6, 10099 Berlin, Germany}
\affil[3]{Singapore Management University, 50 Stamford Road, Singapore 178899}
\affil[*]{Corresponding author: \href{mailto:l.vomfell@warwick.ac.uk}{\texttt{l.vomfell@warwick.ac.uk}}}

\providecommand{\keywords}[1]{\textit{Keywords:} #1}

\date{}

\newcommand\blfootnote[1]{%
	\begingroup
	\renewcommand\thefootnote{}\footnote{#1}%
	\addtocounter{footnote}{-1}%
	\endgroup
}

\begin{document}
	
\maketitle

	\begin{abstract}
    Crime prediction is crucial to criminal justice decision makers and efforts to prevent crime. The paper evaluates the explanatory and predictive value of human activity patterns derived from taxi trip, Twitter and Foursquare data. Analysis of a six-month period of crime data for New York City shows that these data sources improve predictive accuracy for property crime by 19\% compared to using only demographic data. This effect is strongest when the novel features are used together, yielding new insights into crime prediction. Notably and in line with social disorganization theory, the novel features cannot improve predictions for violent crimes.

	\end{abstract}

\blfootnote{This is the peer reviewed version of the following article: L.\ Vomfell, W.K.\ H\"ardle and S.\ Lessman, ``Improving Crime Count Forecasts Using Twitter and Taxi Data'', \textit{Decision Support Systems}, vol.\ 113, pp.\ 73--85, Sep 2018, which has been published in final form at \url{https://doi.org/10.1016/j.dss.2018.07.003}.}

	\section{Introduction}
    Every day, people leave their neighbourhood to commute to work, shop in malls or relax in museums and bars. Such travel creates a social flow of both crime targets and perpetrators that connect areas beyond spatial distance and facilitates criminal activity \citep{wikstrom2010activity}. 
    Exploitation of location-based data offers new perspectives on the mechanisms of crime emergence and helps predict the occurrence of crime. Government institutions and especially police depend on adaptive, short-term crime predictions to anticipate changes and breaks in crime patterns and allocate scarce resources efficiently \citep[e.g.,][]{xue2006spatial}. 
	
    The objective of this paper is to establish the performance of data on human dynamics in predicting crime. In pursuing this goal, the paper proposes predictive models that extend conventional crime forecasts by incorporating three sources of data: public venues, social media activity and taxi flows. We suggest alternative ways to extract features from the data sources and examine how their interaction improves predictive performance. This provides concrete guidance to decision makers on how to leverage these new data sources for accurate crime forecasting. 
        
    Empirical results using crime data from New York City confirm the relevance of the proposed features. Using a rolling-window prediction approach, we demonstrate that including the novel features significantly improves crime predictions for some types of crime. The results reveal interaction effects: Features from different data sources work best when used in combination. 
    Our dual approach of prediction and explanatory analysis addresses policy makers' concerns about preventing crime in a predictive policing and a wider prevention context. In line with social disorganisation and opportunity theory, our results add to a better understanding of the link between crime opportunities and human dynamics and highlight new areas for policy design. 
    
	The paper is organised as follows: Section~\ref{sec:relatedwork} discusses related work. Section~\ref{sec:background} introduces spatial and non-spatial prediction models. Section~\ref{sec:approach} outlines the data sources and feature construction methods. Empirical results are presented in Section~\ref{sec:results} and discussed in Section~\ref{sec:discussion}. Section~\ref{sec:conclusion} concludes the paper.

	\section{Related Work}
	\label{sec:relatedwork}
    Our study uses online data together with spatial analysis to understand behavioural aspects of the emergence of crime and how this improves crime prediction.   
   In this section, we briefly introduce seminal explanatory studies that use spatial analysis to provide empirical support for prominent crime theories. Then, we elaborate how online data sources have been used in explanatory contexts before presenting forecasting studies that use online data or spatial analysis to predict crime.
    
	The main theories concerned with explaining the spatio-ecological dimension of crime are opportunity theory and social disorganisation theory. The former analyses crime events as opportunities created by the intersection of a suitable target, a motivated offender, and lack of supervision \citep{cohen1979social}. 
    Social disorganisation theory considers neighbourhood characteristics that influence the likelihood of criminal activity among inhabitants. A lack of social control and social cohesion within a community combined with structural disadvantages gives rise to criminal behaviour \citet{kubrin2003new}.
    
    Classical crime modelling draws on these theories and uses regression analysis to identify socio-economic predictors of criminal behaviours on an aggregated level. The findings emphasise the relevance of demographic characteristics such as residential instability, ethnic heterogeneity and population density \citep[\eg][]{sampson1997neighborhoods}. These results have been supplemented with spatial analysis to evaluate the relevance of spatial dependence. Spatial proximity to violence has been shown to be more important than demographic data \citep{morenoff2001neighborhood,kubrin2003structural}.
   
    With the availability of online data, crime modelling has shifted to incorporating aggregated, anonymous human behavioural data. Geo-tagged Twitter data in particular has been used to understand how topics on social media relate to crime, for example through term-frequency analysis \citep{williams2017crime}. Dynamic data on human activity has also been used to model crime. \citet{traunmueller2014mining} examine correlations between people activity features, which they derive from mobile phone data, and monthly crime rates. 
    
    Paralleling the development in classical crime modelling, online data has been combined with spatial analysis to explore their relationship. \citet{bendler2014crime} include Twitter and local points of interest (POI) data in a geographically weighted regression to capture human activity and explore spatial dependence between crime locations. They show that only some crimes such as burglary are related to Twitter activity.    
	\citet{wang2016crime} also consider POI data, which they integrate with taxi flow data to model yearly crime rates in Chicago. They find a model using both types of information to outperform models using only POI or taxi data. Such synergy hints at an interdependence between the two sources, which has also been observed by \citet{bendler2014crime}. 
    
    	\afterpage{%
		\clearpage
		\begin{landscape}
			\centering
			\singlespacing
			\renewcommand{\arraystretch}{1.5}
			\begin{tabular}{lccccP{3cm}ll} \toprule
				Study & \makecell[l]{Explanatory/\\predictive}  & \multicolumn{1}{l}{Spatial} &  \makecell[l]{Human\\Dynamics} & \makecell{Machine\\Learning} & Crime Type & City & \makecell[l]{Time\\Frame} \\ \midrule
				\citet{wang2016crime} &E & \checkmark & \checkmark & & all crime & Chicago & yearly \\
				\citet{bendler2014crime} &E& \checkmark & \checkmark & & assault, burglary, homicide, theft, \ldots & San Francisco & hourly\\
				\citet{traunmueller2014mining} & E & & \checkmark & & street vs. indoor & London & monthly\\
				\citet{williams2017crime} & E& & \checkmark & & burglary, theft, drugs, violent crime, \ldots & London & monthly \\
                \citet{gerber2014predicting} & P & & \checkmark & & theft, battery, drugs, burglary, \ldots & Chicago & daily \\
				\citet{xue2006spatial} & P& \checkmark &  & \checkmark & burglary & Richmond, VA & monthly \\
				\citet{rosser2017predictive} & P & \checkmark & & \checkmark & residential burglary & anonymous UK city & daily \\
				\citet{bogomolov2014once} & P & & \checkmark & \checkmark & (hotspot classification) & London & monthly\\
				\citet{kang2017prediction} & P & \checkmark & & \checkmark & all crime & Chicago & daily \\
				\citet{aghababaei2018mining} & P& &  \checkmark&  \checkmark & theft, drugs, burglary, ... & Chicago & daily \\
				\midrule
				\textit{This study} & E and P & \checkmark & \checkmark & \checkmark & violent and property crime & New York & weekly \\
				\bottomrule
			\end{tabular}
			\captionof{table}{Literature Overview}
			\label{tbl:literatureoverview}
		\end{landscape}
		\clearpage
	}

	In contrast to explanatory studies, crime prediction has paid comparatively less attention to the intersection of space and human dynamics and usually uses online data as inputs for crime prediction. For example, \citet{aghababaei2018mining} employ temporal topic detection to identify Twitter topics predicting crime. \citet{bogomolov2014once} train a Random Forest to predict high-crime areas using features related to visitors volumes based on telecommunication records. \citet{gerber2014predicting} find that prediction models using Twitter topic modelling outperform Kernel density estimation-based models.
    
    There are few predictive studies using spatial analysis. Most notably, the work by \citet{rosser2017predictive} analyses criminal incidents on a street segment-level instead of a grid- or census unit-level. However, human dynamics are not explicitly taken into consideration since the predictions are not based on any analysis of traffic volume or pedestrian density on those streets.	
	\citet{xue2006spatial} model the coordinates of crime as a locally optimal site picked by the offender from a set of spatial alternatives to commit the crime. Similar to \citet{rosser2017predictive}, they do not take human dynamics into account.
	An interesting approach to synthesising different data sources for crime prediction is proposed by \citet{kang2017prediction} who train a deep neural network (DNN) to integrate Google Streetview images and temporal features into a joint feature as input layers for DNN-based crime prediction.

	Table~\ref{tbl:literatureoverview} shows how explanatory studies frequently incorporate behavioural data and spatial dependence, whereas predictive studies focus on only one of the two aspects. Therefore, a contribution of this paper is the joint consideration of data on spatial structure and human dynamics. 
    A second contribution stems from combining explanatory and predictive analysis as it is crucial to understand the underlying process of crime generation to not only successfully predict crime incidents but also prevent crime \citep{Camacho-Collados}. 

	\section{Methodology} 
	\label{sec:background}
    Crime rates depend on the underlying population at risk, which need not correspond to the residential population in a geographic unit \citep[eg][]{malleson2015impact}. Therefore, a common modelling approach, which we adopt in this study, is to use counts of crime incidents. Our data forms a panel of crime counts and covariates for 1974 census tracts for 26 weeks, indexed by $i$ and $t$, respectively.
    
    Our analysis approach is two-fold: our main focus is crime prediction, which we supplement with explanatory analysis. We use spatial econometric models and machine learning techniques to fit models and predict crime.       
	In the following section, we first describe the econometric models in Subsections~\ref{sec3:linear} and \ref{sec3:count}. Then, we describe the machine learning methods used in Subsection~\ref{sec3:machine}. We use a rolling window prediction approach which we explain in Subsection~\ref{sec3:predictors} where we also present the linear predictors.    
Before detailing the models, we introduce some notation. 
Modelling crime counts in a city begins with a specific, bounded two-dimensional area $D \subset \mathbb{R}^2$, where $D$ denotes the surface area of the city. $D$ can be partitioned into a finite number $N$ of well-defined, non-overlapping areal units, e.g. census tracts. 

Crime events are modelled as realisations of a point process on $D$. The locations of $k_t$ crime events at time $t$ are denoted by ${S_t = \{s_{1t}, \ldots, s_{k_t t}}\}$. This allows modelling the number of realised events in an areal unit as a time-dependent count variable. Let this count variable be defined as $m(i,t) = \sum_{l = 1}^{k_t}\mathbbm{1}(s_{lt} \in i), i = 1, \ldots, N$ such that $m(i,t)$ gives the number of crimes in unit $i$ at time $t$. Let $y$ denote the vector of $NT$ count variables observed at the $N$ areal units in $T$ periods such that $m(i,t) \equiv y_{it}$. 	

	Spatial dependence between areas can take the form of a Markov random field, which defines a neighbourhood for each element in $y$. An areal unit $j$ is a neighbour of areal unit $i$ if the conditional distribution of $y_i$ depends on $y_j$ \citep{cressie1993statistics}. Let $A_i = \{j : j \text{ is a neighbour of } i\}$ be the neighbourhood of unit $i$. Note that $A_i$ excludes unit $i$. \label{page:markov}
	
	\subsection{Linear Models}
	\label{sec3:linear}
	Consider the simple pooled linear panel regression model:
	\begin{equation}
	y = X\beta + e, \quad e \sim N(0, \sigma^2I_{NT}),
	\label{eq3:linear}
	\end{equation}
	where $X$ is a $NT\times K$ matrix of $K$ regressors.
	In the presence of spatial dependence, the error terms in \eqref{eq3:linear} are no longer uncorrelated. Approaches to account for such error correlation include the simultaneous autoregressive (SAR) and the conditional autoregressive (CAR) model.
	
	The SAR model introduces spatial structure through a spatial lag \citep[p.~406]{cressie1993statistics}:
	\begin{equation}
	y = (I_T \kron \rho W)y + X\beta + \varepsilon, \quad \varepsilon \sim N(0,\sigma^2I_{NT}), \label{eq3:sar}
	\end{equation}
	where $\kron$ denotes the Kronecker product, $I_T$ denotes the identity matrix of order $T$, and $W$ is a $N \times N$ binary matrix specifying which areas are spatially adjacent with $w_{ii} = 0 \;\forall i$. $\rho$ is the parameter that specifies the magnitude of spatial dependence. 
	
	The inclusion of a spatial lag of the dependent variable accounts for spatial spillovers and a mismatch of the spatial scale with the spatial event. Both effects occur in crime modelling since the contagion effect of crimes leads to a diffusion through space. In addition, economic and criminal features do not match perfectly with the spatial units. A spatial lag SAR model is a convenient choice to account for these characteristics \citep{anselin2008spatial}.
	
	The CAR model introduces a spatial dependence parameter in the error term which accounts for small-scale spatial variation \citep[p.~407]{cressie1993statistics}.	
	 This yields the following model:
	\begin{IEEEeqnarray}{rCl}
		y &=& X\beta \> + \varepsilon, \label{eq3:CAR}\\
		\varepsilon &\sim& N\left(0, \sigma^2\{I_T \kron (I_N-\delta W)^{{-1}}\}\right), \nonumber
	\end{IEEEeqnarray}
	where $W$ is again a $N \times N$ spatial adjacency matrix and $\delta$ denotes the magnitude of spatial dependence between neighbouring regions. 
    
    The CAR model introduces spatial structure as a Markov random field, such that the conditional distribution of each area depends on the neighbourhood.
	The distribution of $y_{it}$ conditional on all $y_{jt}$ can be shown to be
	\begin{IEEEeqnarray}{r}
		y_{it} | y_{jt} \sim N\left(X_{it}^\top\beta + \sum_{j} \delta W_{ij} (y_{jt} - X_{jt}^\top\beta), \sigma_i^2\right),
	\end{IEEEeqnarray}
	for $i \neq j$, where $\sigma^2_i$ denotes the conditional variance \citep[p.~407]{cressie1993statistics}. 
    This conditional dependence structure is different from the structure modelled in a SAR model. There, the inclusion of the spatial lag means that values in unit $i$ do not only depend on values in the direct neighbourhood $A_i$ but also on higher-order neighbours, \ie neighbours of neighbours. Therefore, the SAR model implies a global dependence structure compared to the CAR model \citep{anselin2008spatial} 
	 	
	\subsection{Count Models}
	\label{sec3:count}
	Linear models offer a broad framework to include spatial structure but fail to accommodate the integer-valued and non-negative nature of crime counts. Small counts are better modelled by a Poisson Generalised Linear Model.
	In the case of crime counts, the Poisson parameter $\lambda$ represents the expected incident count: 
	\begin{IEEEeqnarray}{r}
		\lambda = E(y\,|\,X) = e^{X^\top\beta}.
		\label{eq3:poissonNew}
	\end{IEEEeqnarray}
	
    Similar to the linear model, the errors of the Poisson model in \ref{eq3:poissonNew} are no longer uncorrelated under spatial dependence. Poisson Generalised Linear Mixed Models (GLMMs) account for this dependence by incorporating a random effect in the GLM predictor. GLMMs model $ E(y\,|\,X)$ as a linear combination of fixed effects $X$ and random effects $Z$ with a logarithmic link function \citep{agresti2007introduction}:
	\begin{equation}
	\log \lambda_{it} = X_{it}^\top \beta + Z_{it}\eta_i \label{eq3:poissonCAR}.
	\end{equation}
	Here, $Z\eta$ are location-specific random effects. At each cross-section $t$, $Z$ is a $N \times N$ indicator matrix of the spatial units, which means that the random effect is simply a random intercept added to the conditional mean.
	The distribution of the random vector $\eta$ is assumed to be multivariate normal: 
	\begin{equation}
	\eta \sim N(0, D), \quad D = \sigma^2Q^{-1}. \label{eq3:sigma}
	\end{equation}
	$Q$ is a symmetric spatial dependency matrix different from the adjacency matrix $W$ used before. Its entries are as follows: 
	\begin{equation}
	Q_{ij} = 
	\begin{cases}
	\vert A_i \vert & \text{if } i = j,\\
	-1 & \text{if } j \in A_i \text{ and } i \neq j,\\
	0 & \text{if } j \notin A_i \text{ and } i \neq j,\\
	\end{cases} \label{eq3:Q}
	\end{equation}
	where the $\vert A_i\vert$ entries on the diagonal denote the size of the neighbour set and neighbours are indicated by $-1$ \citep[p.~186]{leroux2000estimation}. 
	In the non-spatial Poisson GLM in \eqref{eq3:poissonCAR}, the variance is equal to the expectation \citep{agresti2007introduction}. In the model in \eqref{eq3:poissonCAR}, this is not the case. Here, $\sigma$ accounts for both the variance and spatial dependence.	
	The parameters in \eqref{eq3:poissonCAR} and \eqref{eq3:sigma} are estimated using restricted maximum likelihood (REML) and Fisher Scoring \citep{kneibrestricted}.
	
    	\subsection{Machine Learning Models}
	\label{sec3:machine}
	Previous models make assumptions about the data-generating process and consider a linear additive relationship between crime counts and covariates. Machine learning techniques are more flexible and account for non-linearity in a data-driven manner \citep{Kuzey}. We concentrate on random forest (RF), gradient boosting machines (GBMs), and feed-forward artificial neural networks (ANNs), all of which have shown promising results in previous studies \citep[\eg][]{Bhattacharyya,Delen}. 
	
	RF develops an ensemble of size $k$ through drawing $k$ bootstrap samples from the training data. The base models in RF consist of individual decision trees, which are grown from the bootstrap samples. To increase randomness among the base models, RF determines the best split during tree growing among a randomly sampled subset of covariates \citep{breiman2001random}. The model prediction consists of the simple average calculated across the $k$ base models.
	
	GBMs embody the idea of additive modelling. The algorithm incrementally develops an ensemble through adding base models. In our paper, we use regression trees as base models. These are fitted to the residuals via the negative gradient of the loss function of the current ensemble. GBM predictions are obtained by calculating a weighted average over base model forecasts, whereby the weights are determined during gradient descent \citep{Friedman_SGB}. 
	
	An ANN model consists of interconnected layers of processing units (neurons) with connection weights representing the model parameters. Estimating an ANN model involves minimising loss functions with respect to connection weights using gradient-based methods. ANNs calculate the output of a neuron as a non-linear transformation of the weighted sum over its input neurons. The transformations are called activation functions and allow an ANN to capture non-linear patterns in data \citep{Kim_DSS}. We use a Rectified Linear Unit (ReLU) activation function.
    
	\subsection{Rolling window prediction}
	\label{sec3:predictors}
            We use a rolling window prediction approach where we use all $y_{1:t} = (y_{1, 1:t}, \ldots, y_{N, 1:t})^\top$ to estimate our models and produce forecasts $\hat{y}_{t+1} = (\hat{y}_{1, t+1}, \ldots, \hat{y}_{N, t+1})^\top$ for the next week. We compute the prediction errors $e_{t+1} = y_{t+1} -\hat{y}_{t+1}$. We repeat this step for $t = h, \ldots, T - 1$ where $h$ is the smallest number of observations used for estimating the model. We set $h = T/2$. We then calculate the total mean squared error based on the obtained errors $MSE = \frac{1}{N h}\sum_{i=1}^N \sum_{t = h+1}^{T} e_{it}^2$. 
            
	For the linear model, the predictions for weekly crime counts are obtained by using the best linear unbiased predictor or its panel equivalent \citep{baltagi2011estimating}. Table~\ref{tbl3:predictors} gives the predictors for the time period $t+1$ for the regression models.
	\begin{table}[ht]
		\centering
		\renewcommand{\arraystretch}{1.25}
		\begin{tabular}{lr@{\hskip 4pt}c@{\hskip 4pt}l} \toprule
			Model & \multicolumn{3}{l}{Predictor} \\ \midrule
			LR & $\hat{y}_{t+1}$ &=& $X_{t+1}\hat{\beta}$ \\
			SAR & $\hat{y}_{t+1}$ &=& $(I_N - \rho W)^{-1}X_{t+1}\hat{\beta}$ \\
			&&& $+\> (I_N - \rho W)^{-1}\hat{\varepsilon}$ \\
			CAR & $\hat{y}_{i, t+1}$ &=& $X_{i,t+1}^\top\hat{\beta} + \sum_j\delta w_{ij} $ \\
			& &&$ \left((1/t)\sum_{k=1}^t (y_{jk} - X_{jk}^\top\hat{\beta}) \right)$ \\
			GLM & $\hat{y}_{t+1}$ &=& $\exp(X_{t+1}\hat{\beta}) $\\
			GLMM &$ \hat{y}_{t+1}$ &=& $\exp(X_{t+1}\hat{\beta} + Z_{t+1}\hat{\eta})$ \\ \bottomrule
		\end{tabular}
		\caption{Predictors for the spatial linear regression models considered in the study.} 
		\label{tbl3:predictors}
	\end{table}
	The SAR predictor is obtained by spatially lagging the linear predictor and adding the spatially lagged error vector of the model. The CAR predictor is obtained by taking a time-averaged conditional expectation.
    
	Machine learning models require auxiliary data for hyperparameter tuning to enable adaption of a learning algorithm to a given task \citep[e.g.,][]{Carneiro.2017}. For such models, we use the first $1, \ldots, t - 2$ weeks in the window of length $t$ as training set and the last two weeks as validation set for parameter tuning. This way, we still produce an out-of-sample one-step ahead forecast for $t+1$. We report the models with the lowest prediction errors on the test set. We tune the hyperparameters using grid search (see \ref{ap:grid} for details) at each window. Since we include lagged crime counts as predictors, the rolling window approach corresponds to cross-validation for time-dependent data.

	\section{Data Integration and Feature Construction}
	\label{sec:approach}
    Since the data sources we use (census, POI, Twitter and taxi flow data) have different time coverages, we use the most recent complete overlap from June 1, 2015 to November 29, 2015. We aggregate the temporal data to weekly intervals which begin uniformly on Monday. The final data set covers 26 weeks. As discussed in Section~\ref{sec3:predictors}, we set $h = T/2 = 13$ weeks. This choice results in 13 windows of length $1:t, t = 13, \ldots, 25$, on which we train our models. We then produce 13 separate one-step ahead forecasts for week $t+1$. We use the human dynamics features at time $t$ to predict crime counts at time $t+1$.
    
    The short time frame of the data makes explicit modelling of temporal effects infeasible since 26 weeks are not sufficient to reliably estimate weekly or monthly seasonality.     
    We also do not include a dummy for the week of the first of a month to account for a potential ``pay day effect'': While one might expect that criminal behaviour associated with drinking increases after receiving the monthly salary, this implied human activity is already captured by our novel data sources.

	The following subsections introduce the data sources. For each source, we elaborate on alternative options for feature engineering since different formulations may differ in their predictive power. The definitions always include a general definition using raw counts and additional versions similar to data standardisation or variance reduction such a log-transformed features. We do not consider further feature transformations such as Principal Component Analysis due to their non-interpretability. Section~\ref{sec4:eval} details how the final set of features has been selected. 
	
	\subsection{Census}
    	The spatial units of analysis are census tracts as defined by the US Census Bureau. We use the coordinates of point-referenced data to match them to the corresponding census tract.
    We select the following eight demographic variable from Summary File 1 of the 2010 census data \citep{census} based on previous studies \citep[\eg,][]{wang2016crime}: the total population in the census tract, the median age of the population, the share of males, the share of the Black, Asian, and Hispanic population, respectively, the rate of female-headed family households, and the rate of vacant accommodation. 
	
	\subsection{New York City Crime Data}
	\label{sec4:data}
	Data on criminal incidents is provided by the New York City Police Department \citep{nypd}. We focus on violent and property crime because their spatial distribution differs, which facilitates examining the proposed features in a context of varying spatial dependence. Violent crime encompasses murder and non-negligent manslaughter, robbery, and aggravated assault. Since rape incidences are not geo-located in the NYPD dataset, we exclude them from the analysis. Property crime comprises burglary, larceny-theft, motor vehicle theft, and arson. 
	
\begin{figure}
		\centering
		\begin{subfigure}{0.4\textwidth}
			\centering
			\includegraphics[width=0.8\linewidth]{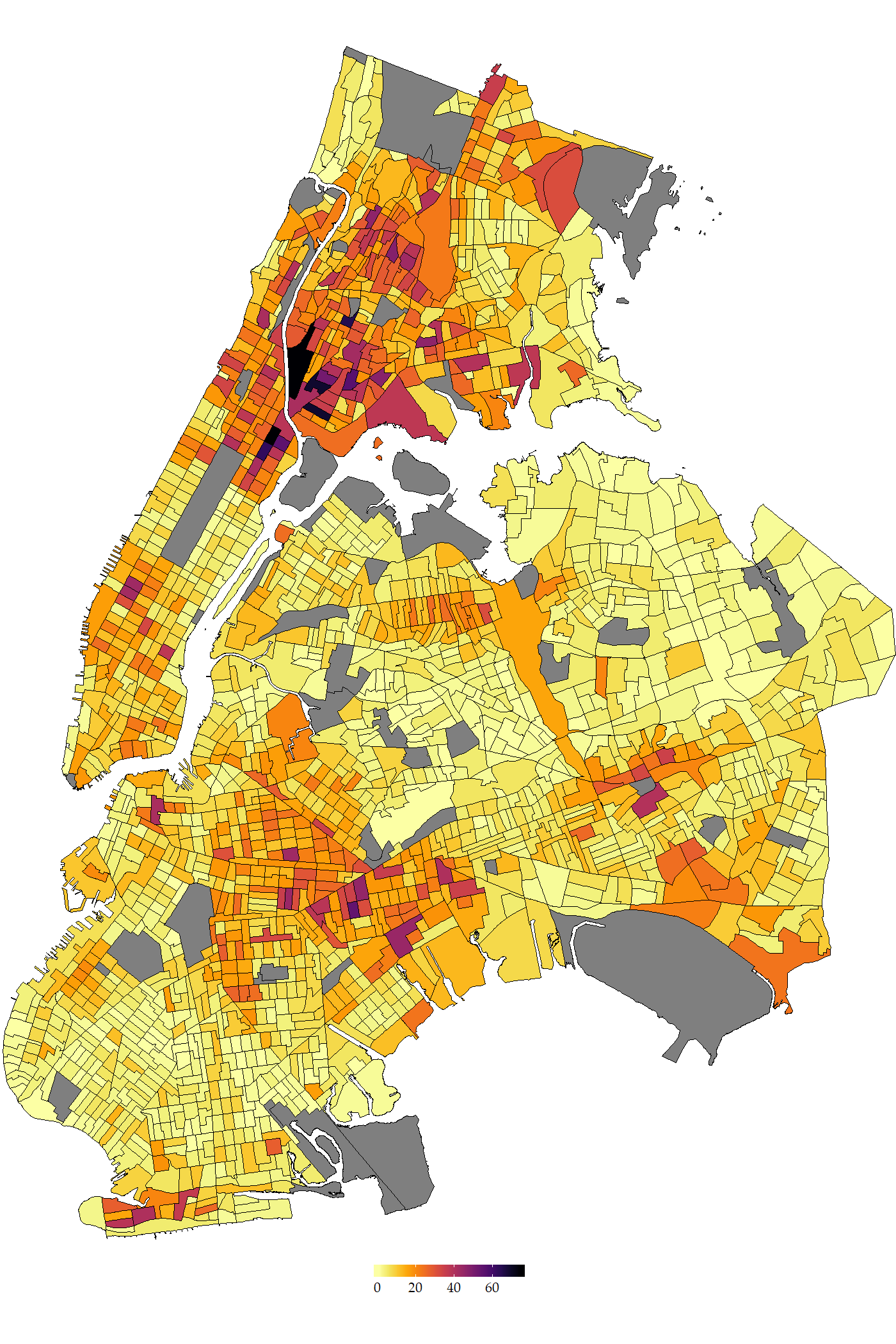}
			\subcaption{Violent Crime}
			\label{fig4:violentmap}
		\end{subfigure}\quad
		\begin{subfigure}{0.4\textwidth}
			\centering
			\includegraphics[width=0.8\linewidth]{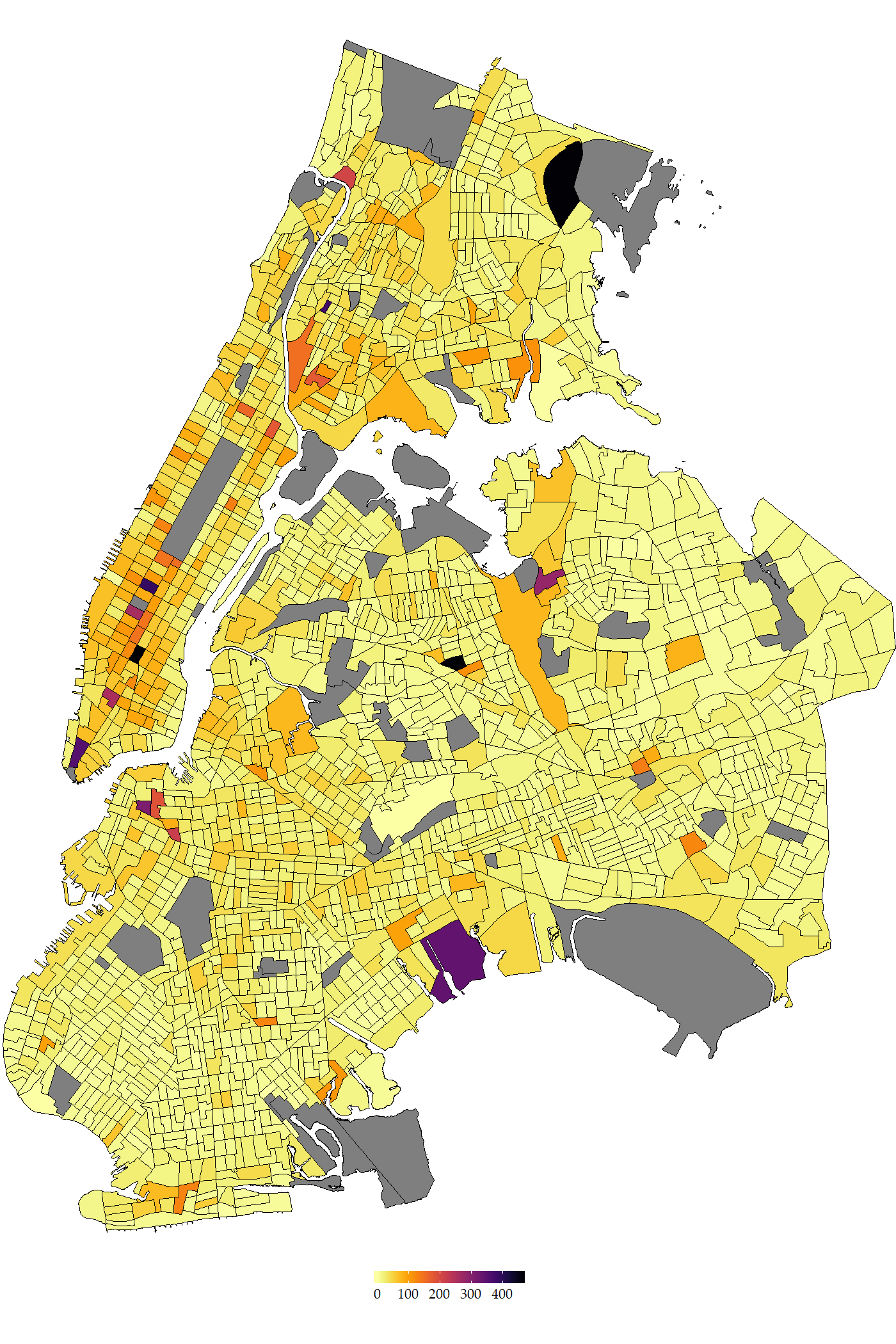}
			\caption{Property Crime}
			\label{fig4:propertymap}
		\end{subfigure}
		\caption{Number of crime incidents between June and November 2015. In the property crime map, the area around Penn Station (largest outlier with 2002 incidents) is excluded for more consistent colour scaling.}
		\label{fig4:crimemap}
	\end{figure}

	Figures \ref{fig4:violentmap} and \ref{fig4:propertymap} show the spatial distribution of crime for the analysis period of June to November 2015. Property crime exhibits a more even distribution than violent crime.
	The strength of spatial correlation between areas is tested using Moran's $I$ \citep{anselin2008spatial}. For both crime types and every time period, the null hypothesis of no spatial dependence is rejected with $p < 0.000$.
	
	\subsection{Foursquare} \label{sec4:foursquare}
	We gather POI data from Foursquare, a mobile recommendation app. We consider POI data a characterisation of the census tract since POI categories attract specific groups of people. For example, one can expect that more nightlife venues attract drunken behaviour. Prior work has evidenced a connection between criminal activity and local points of interest in a geographic area \citep{bendler2014crime}. Foursquare categorises all venues along nine main dimensions: nightlife, food, arts \& entertainment, residence, shops, travel, outdoors \& recreation, college \& education, and professional.
	In total, we obtain 47,113 POI in the geographic area of interest. 
	
	Two different ways of constructing the feature from POI data are considered: 
        \begin{enumerate*} \item the total counts of venues per category, \item the share of categories on the total number of venues in the census tract.	\end{enumerate*}
	
	\subsection{Taxi}
	\label{sec4:taxi}
\begin{figure}
		\centering
		\begin{subfigure}{0.4\textwidth}
			\centering
			\includegraphics[width=0.8\linewidth]{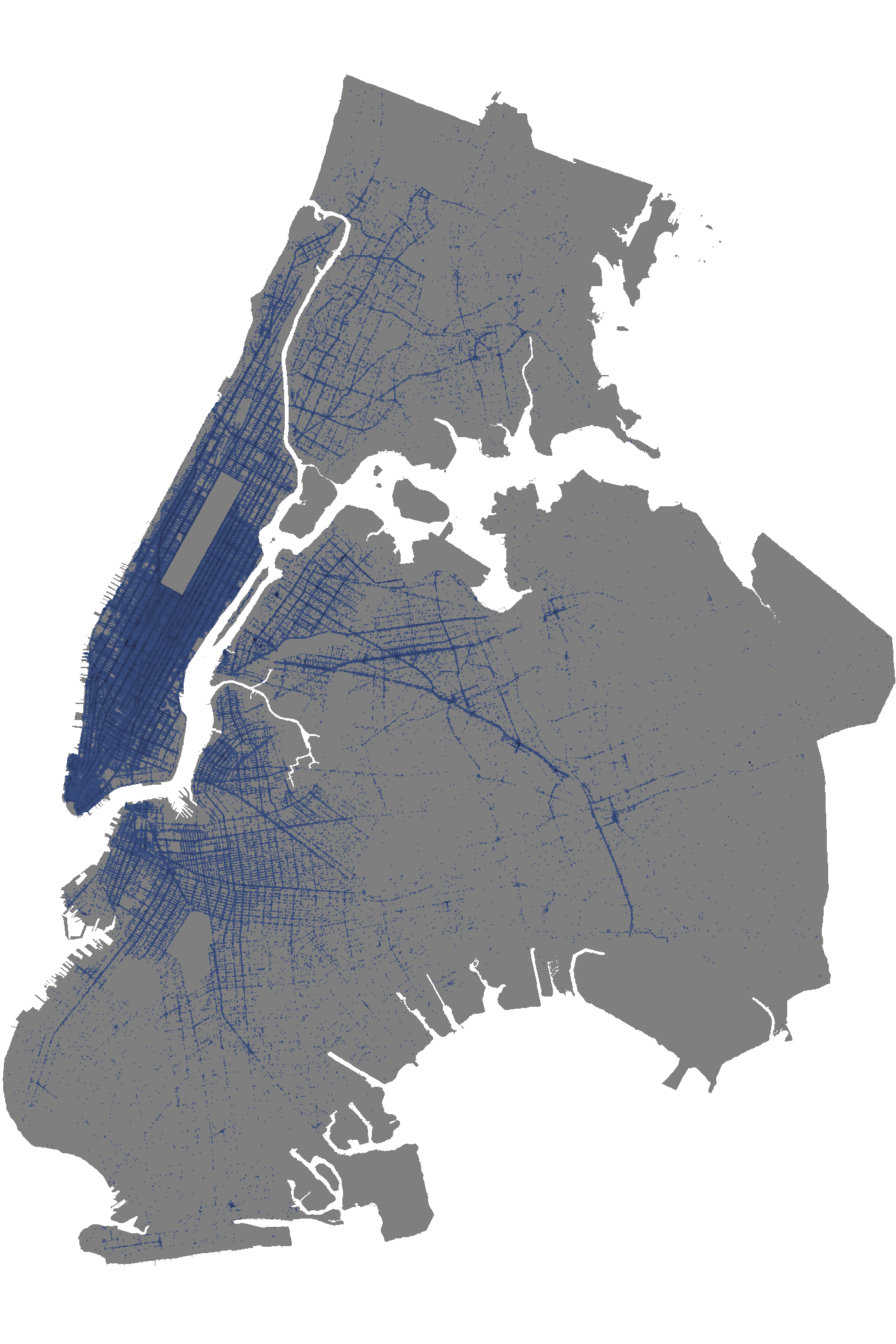}
			\subcaption{Pickups}
			\label{fig4:taxipickups}
		\end{subfigure}\quad
		\begin{subfigure}{0.4\textwidth}
			\centering
			\includegraphics[width=0.8\linewidth]{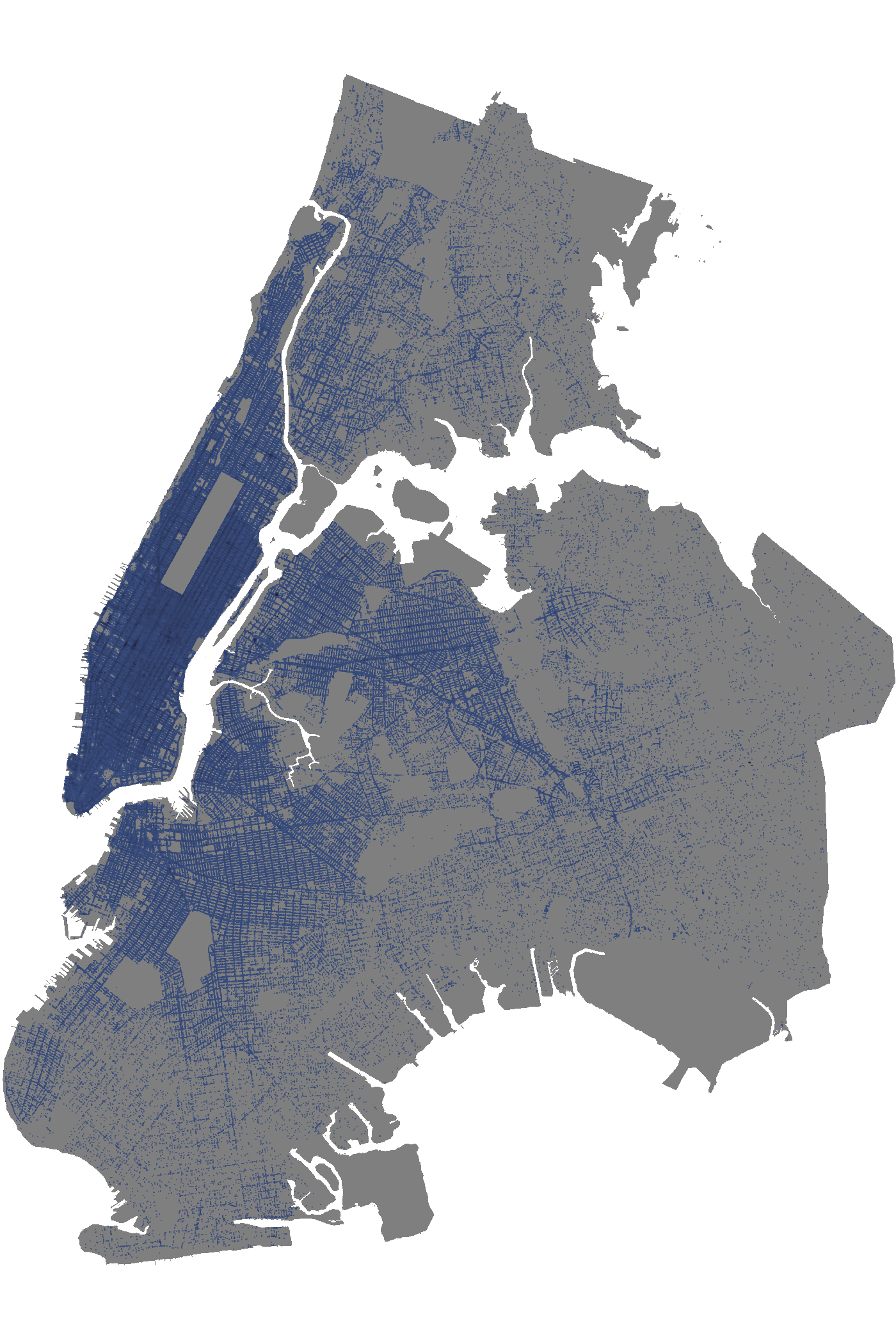}
			\caption{Dropoffs}
			\label{fig4:taxidropoffs}
		\end{subfigure}
		\caption{Coordinates of complete taxi trips in New York City in week 46 in 2015.}
		\label{fig4:taxi}
	\end{figure}

	The \citet{TLC} provides taxi flow data. We argue that taxi flows illustrate connections between different neighbourhoods beyond what is already covered through spatial proximity. Around 25\% of all taxi trips end in a census tract that is not a neighbour of the tract they started in, suggesting that the taxi feature captures connections between census tracts that go beyond spatial proximity. Figure~\ref{fig4:taxi} supports this view and, in agreement with \citet{wang2016crime}, confirms taxi data as a valuable source for crime modelling.
    
    We consider all trips within New York City in the analysis time frame but exclude trips that start or end outside the analysis area. This gives 70,288,218 trips in the 26 weeks. We aggregate individual trips to a weekly connection flow matrix $F$, with rows (columns) of $F$ referring to the census tract where the trip started (ended). Hence, $f_{ij}$ denotes the number of trips made from tract $i$ to $j$ for each time interval. Note that $f_{ii} = 0\,\forall i$ as otherwise, crime rate of census tract $i$ would be used as its own predictor. 
	
	The taxi flow feature is then constructed as $c_t = F_t y_{t-1}$ such that neighbouring crime rates are weighted by the magnitude of flow $F$.
	It is crucial to note that the crime vector $y$ is lagged by a week to prevent unintended implicit simultaneity of the response $y_t$ and its predictors. The week index $t$ is dropped for ease of notation.
	
	We propose three  different ways to construct $c$ and demonstrate the calculation for $c_1$, the feature of example tract 1:
	{ \setlength\abovedisplayskip{2pt}
		\setlength\belowdisplayskip{2pt}
		\begin{enumerate}
			\item \label{taxi1} Raw multiplication: One can define $t$ as the simple matrix multiplication of the flow matrix $F$ and the crime count vector $y$: 
			\begin{equation}
			c_1 = f_{12}y_2 + \ldots + f_{1N}y_N. \nonumber
			\end{equation}	
			\item Normalised by source: The taxi flow arriving in each census tract is normalised by the total number of flows leaving the source census tract. For example, the flow leaving the second census tract towards the first tract is normalised by all flows leaving from the second tract:
			\begin{equation}
			c_1 = \frac{f_{21}}{\raisebox{-1pt}{$f_{21} + f_{23} + \ldots + f_{2N}$}} y_2 + \ldots + \frac{f_{N1}}{\raisebox{-1pt}{$\sum_{i = 1}^{N} f_{Ni}$}} y_N. \nonumber
			\end{equation}
			\item Normalised by destination: 
            The taxi flow arriving in each census tract is normalised by the total number of flows arriving in the destination census tract:
			\begin{equation}
			c_1 = \frac{f_{21}}{\raisebox{-1pt}{$f_{21}+ f_{31} + \ldots + f_{N1}$}} y_2 + \ldots + \frac{f_{N1}}{\raisebox{-1pt}{$\sum_{i=1}^{N}f_{i1}$}}y_N.  \nonumber
			\end{equation}
		\end{enumerate}
	}
	
	\subsection{Twitter}
    We use Twitter data as a proxy for day-to-day population density through tourists or visitors. Accordingly, we focus on the number of Tweets in an area but do not attempt to extract their topical content. While Foursquare data covers venues as potential destinations of human activity and taxi flow data records where people move to, some of the overall activity is not captured. For example, we observe high numbers of tweets in the census tract containing the 9/11 Memorial site, unaccounted for by any other feature, whether novel or demographic.
   
   We source Twitter data from \citet{twitterdata} who provide IDs to tweets published in the United States between June 1, 2015 and November 30, 2015. We aggregate the number of tweets per week and census tract, and implement four versions  of the Twitter feature: \begin{enumerate*} \item Using the full activity, \item counting night-time tweets only, \item using log-transformed full activity, \item using log-transformed night-time activity \end{enumerate*}. 
	Any tweet sent out between 22pm and 6am contributed to the night-time feature. Taking the logarithm of the number of tweets serves to reduce variation between census tracts. 
    
	\subsection{Evaluation and Feature Selection}
	\label{sec4:eval}
	We proposed multiple variable definitions for each novel data source. Since we are interested in interactions, we select the best combination of all feature types using a variable selection procedure where we estimate CAR models for all possible combinations of the definitions. We then produce one-step ahead forecasts for 13 weeks in total using the procedure described in Section~\ref{sec3:predictors} and pick the combination with the overall lowest MSE. In comparison with in-sample goodness-of-fit statistics such as $R^2$, the MSE-based selection strategy emphasises the predictive value of a feature on out-of-sample data. We suggest that a prediction-centric feature selection strategy is better aligned with the goal of forecasting crime accurately.

    While some machine learning techniques such as Random Forests entail variable importance rankings that can guide variable selection, they may pick up non-linear relationships that linear models cannot accommodate. This would give machine learning models an advantage in subsequent comparisons. To counterbalance this, we select feature definitions through optimising predictions of a linear model. Out of the linear models, we choose the CAR model because it models the outcome variable as a linear combination directly (rather than on the log scale) and because the implied spatial dependence structure is local rather than global. Therefore, the selected feature definition combination is expected to suit the wide range of spatial and non-spatial models we consider.

	\begin{table}
		\begin{subtable}{.5\linewidth}
			\centering
			\begin{tabular}{p{1.4cm}@{ }S[table-format=1.4,detect-all]@{ }S[table-format=1.4,detect-all]@{ }S[table-format=1.4,detect-all]} \toprule 
				Twitter& \multicolumn{3}{c}{Taxi}\\ \cmidrule{2-4}
				& {Raw} & {Destination} & {Source} \\ \midrule
				All & 4.5415 & 4.5221 & 4.8355 \\
				Night & 4.5301 & 4.5272 & 4.8033 \\
				log All & 4.5259 & 4.5329 & 4.8744 \\
				log Night & 4.5193 & $\mathbf{4.5051}$ & 4.8626 \\ \bottomrule
			\end{tabular}
            \caption{Property crime}\label{tbl4:peval}
		\end{subtable}%
		\begin{subtable}{.5\linewidth}
			\centering
			\begin{tabular}{p{1.4cm}@{ }S[table-format=1.4,detect-all]@{ }S[table-format=1.4,detect-all]@{ }S[table-format=1.4,detect-all]} \toprule
				Twitter	& \multicolumn{3}{c}{Taxi}\\ \cmidrule{2-4}
				& {Raw} & {Destination} & {Source} \\ \midrule
				All & 0.5402 & 0.5396 & 0.5400 \\
				Night & 0.5411 & 0.5404 & 0.5400 \\
				log All & 0.5418 & 0.5405 & 0.5414 \\ 
				log Night & 0.5417 & $\mathbf{0.5392}$ & 0.5398 \\ \bottomrule
			\end{tabular}
            \caption{Violent crime}\label{tbl4:veval}
		\end{subtable} 
		\caption{MSE values for crime predictions from CAR models including POI data in the form of the total counts of venues per Foursquare category together with alternative definitions of the Twitter and taxi features.}
	\end{table}
	
		Tables~\ref{tbl4:peval} and \ref{tbl4:veval} show MSE values for property and violent crime. 
	We present results for alternative definitions of the Twitter and taxi features. Total counts of venues for Foursquare category produces uniformly better results than the venue share. 
	Overall, we observe the best results with the non-normalised POI feature, log-transformed nightly tweet activity, and taxi data normalised by destination. For the POI feature, however, using total counts outperforms normalisation. The counts preserve differences in the POI distribution across New York City, which results in better predictions than the shares of categories. 
    
    \begin{table}
	\centering
    \sisetup{table-number-alignment = center, group-separator = {,}, group-four-digits}
	\begin{tabular}{l*{5}{S[table-format = 4.2]}}
		\toprule
		Variable & {Mean} & {Std. deviation} & {Median} & {Min} & {Max} \\ 
		\midrule
		Property crime & 1.45 & 2.34 & 1.00 & 0.00 & 56 \\ 
		Violent crime & 0.37 & 0.74 & 0.00 & 0.00 & 11 \\ 
		Population & 3829.61 & 2118.97 & 3431.50 & 56 & 26588 \\ 
		Median age & 35.92 & 6.01 & 35.40 & 13.40 & 80.90 \\ 
		Male & 0.48 & 0.03 & 0.48 & 0.32 & 0.94 \\ 
		Black & 0.28 & 0.31 & 0.12 & 0.00 & 0.96 \\ 
		Asian & 0.13 & 0.16 & 0.06 & 0.00 & 0.88 \\ 
		Hispanic & 0.27 & 0.23 & 0.18 & 0.00 & 0.91 \\ 
		Vacancy rate & 0.08 & 0.06 & 0.07 & 0.00 & 0.65 \\ 
		Female-headed HH & 0.20 & 0.12 & 0.17 & 0.00 & 0.58 \\ 
		log night tweets & 1.22 & 1.42 & 0.69 & 0.00 & 7.87 \\ 
		Entertainment POI & 2.90 & 3.39 & 2.00 & 0 & 64 \\ 
		Uni POI & 2.51 & 3.43 & 2.00 & 0 & 61 \\ 
		Food POI & 3.01 & 3.01 & 2.00 & 0 & 28 \\ 
		Professional POI & 2.61 & 2.47 & 2.00 & 0 & 20 \\ 
		Nightlife POI & 2.82 & 2.81 & 2.00 & 0 & 27 \\ 
		Outdoors POI & 2.29 & 2.33 & 2.00 & 0 & 19 \\ 
		Shops POI & 2.75 & 2.73 & 2.00 & 0 & 26 \\ 
		Travel POI & 2.52 & 2.64 & 2.00 & 0 & 26 \\ 
		Residential POI & 2.76 & 2.51 & 2.00 & 0 & 22 \\ 
		Taxi (property) & 1.45 & 3.21 & 0.28 & 0.00 & 58.26 \\ 
		Taxi (violent) & 0.37 & 0.77 & 0.08 & 0.00 & 22.92 \\ 
		\midrule
		\multicolumn{6}{l}{$N = 1974$ census units observed over $T=26$ weeks: 51,324 observations} \\
		\bottomrule
	\end{tabular}
\caption{Summary statistics for the data set}\label{tbl:summstat}
\end{table}
    
    We provide a short data overview in Table~\ref{tbl:summstat}. We find that the new features have low correlations with the demographic variables (all Pearson's $r < 0.35$) but higher correlations with crime of up to 0.63. This makes them valuable predictors in addition to the demographic variables which capture characteristics of the residential population only.
	
	\section{Results}
	\label{sec:results}
	We consider eight different combinations of the features to investigate interactions. The census data serves as baseline and is included in all settings. The other groups are added in all possible combinations which we number from 1 to 8 (Table~\ref{tbl4:settings}). 
	
	\begin{table}
		\centering
		\begin{tabular}{lcccccccc} \toprule
			Features& \multicolumn{8}{c}{Settings} \\ \cmidrule(l){2-9}
			& {1} & {2} &{3} & {4} & {5} & {6} & {7} & {8}  \\ \midrule
			Census & {\checkmark} & {\checkmark} & {\checkmark} & {\checkmark} & {\checkmark} & {\checkmark} & {\checkmark} & {\checkmark} \\
			POI &  & {\checkmark} & {\checkmark} & {\checkmark} &  &  &  & {\checkmark} \\
			Taxi &  &  & {\checkmark} &  & {\checkmark} & {\checkmark} &  & {\checkmark} \\
			Twitter &  &  &  & {\checkmark} & {\checkmark} &  & {\checkmark} & {\checkmark} \\ \bottomrule
		\end{tabular}
		\caption{Definition of experimental settings in terms of different groups  of crime predictors}
		\label{tbl4:settings}
	\end{table}
	
	We begin with examining the explanatory power of the individual features and their interactions. In view of the large number of fitted models (2 types of crime $\times$ 5 model specifications $\times$ 8 settings over 13 windows), we do not reproduce all results.     
    Instead, Tables~\ref{tbla:fullproperty} and \ref{tbla:fullviolent} show the regression coefficients only for the largest possible window of 25 weeks and for setting 8, which includes all feature groups. As detailed in \ref{ap:coef}, the coefficients are stable over different fitting windows. 
	
	\begin{table}[h!]
		\small
		\centering
		\sisetup{input-symbols=(), table-number-alignment = center, group-digits=false, table-space-text-post = \sym{\sym{***}}
		} 
		\begin{tabular}{l*{5}{S[table-format=-1.4, table-align-text-post=false]}}
			\toprule
			Variable& {CAR} & {SAR} & {LR} & {GLM\textsuperscript{1}} & {GLMM\textsuperscript{1}} \\ \midrule
Intercept & -0.4255 & -0.9090\sym{***} & -0.9500 & -0.2687\sym{***} & -1.5419\sym{***} \\ 
   & (0.2448) & (0.2220) & (0.2246) & (0.0794) & (0.0001) \\ 
  Population & 0.0001\sym{*}& 0.0001\sym{***} & 0.0001 & 0.0001\sym{***} & 0.0001\sym{***} \\ 
   & (0.0000) & (0.0000) & (0.0000) & (0.0000) & (0.0000) \\ 
  Median age & 0.0124\sym{***} & 0.0011 & -0.0003 & -0.0076\sym{***} & -0.0012\sym{***} \\ 
   & (0.0024) & (0.0018) & (0.0018) & (0.0008) & (0.0003) \\ 
  Male & -1.3028\sym{***} & 0.4165 & 0.8414 & -0.5642\sym{***} & 0.2675\sym{***} \\ 
   & (0.3929) & (0.3708) & (0.3752) & (0.1331) & (0.0000) \\ 
  Black & 0.7076\sym{***} & 0.1839\sym{**} & 0.2690 & 0.3999\sym{***} & 0.7297\sym{***} \\ 
   & (0.0920) & (0.0644) & (0.0652) & (0.0299) & (0.0001) \\ 
  Asian & 0.5009\sym{***} & 0.1802\sym{**} & 0.2511 & 0.1389\sym{***} & 0.2387\sym{***} \\ 
   & (0.1066) & (0.0690) & (0.0698) & (0.0322) & (0.0000) \\ 
  Hispanic & 0.9776\sym{***} & 0.2430\sym{**} & 0.3017 & 0.4787\sym{***} & 0.6541\sym{***} \\ 
   & (0.1029) & (0.0749) & (0.0758) & (0.0339) & (0.0001) \\ 
  Vacancy rate & 2.3155\sym{***} & 2.0637\sym{***} & 2.3373 & 0.6015\sym{***} & 0.8929\sym{***} \\ 
   & (0.2095) & (0.1777) & (0.1799) & (0.0519) & (0.0000) \\ 
  Female-headed HH & -0.1474 & 1.3941\sym{***} & 1.5020 & -0.0366 & -0.6935\sym{***} \\ 
   & (0.2418) & (0.2054) & (0.2078) & (0.0887) & (0.0001) \\ 
  log night tweets & 0.0987\sym{***} & 0.1221\sym{***} & 0.2034 & 0.2344\sym{***} & 0.0682\sym{***} \\ 
   & (0.0099) & (0.0087) & (0.0089) & (0.0031) & (0.0032) \\ 
  Entertainment POI & 0.0151\sym{***} & 0.0157\sym{***} & 0.0171 & -0.0055\sym{***} & -0.0013 \\ 
   & (0.0036) & (0.0036) & (0.0036) & (0.0013) & (0.0013) \\ 
  Uni POI & -0.0025 & 0.0012 & 0.0023 & 0.0000 & 0.0015 \\ 
   & (0.0032) & (0.0032) & (0.0032) & (0.0013) & (0.0016) \\ 
  Food POI & 0.0543\sym{***} & 0.0512\sym{***} & 0.0441 & 0.0218\sym{***} & 0.0293\sym{***} \\ 
   & (0.0045) & (0.0045) & (0.0046) & (0.0017) & (0.0019) \\ 
  Professional POI & 0.0185\sym{***} & 0.0162\sym{**} & 0.0221 & 0.0241\sym{***} & 0.0240\sym{***} \\ 
   & (0.0055) & (0.0055) & (0.0056) & (0.0020) & (0.0023) \\ 
  Nightlife POI & -0.0599\sym{***} & -0.0704\sym{***} & -0.0761 & -0.0334\sym{***} & -0.0141\sym{***} \\ 
   & (0.0049) & (0.0048) & (0.0049) & (0.0017) & (0.0019) \\ 
  Outdoors POI & 0.0222\sym{***} & 0.0114\sym{*}& 0.0157 & 0.0138\sym{***} & 0.0127\sym{***} \\ 
   & (0.0058) & (0.0058) & (0.0058) & (0.0021) & (0.0023) \\ 
  Shops POI & 0.1433\sym{***} & 0.1236\sym{***} & 0.1209 & 0.0581\sym{***} & 0.0552\sym{***} \\ 
   & (0.0049) & (0.0049) & (0.0049) & (0.0017) & (0.0012) \\ 
  Travel POI & 0.0335\sym{***} & 0.0349\sym{***} & 0.0316 & -0.0021 & 0.0124\sym{***} \\ 
   & (0.0051) & (0.0048) & (0.0049) & (0.0017) & (0.0019) \\ 
  Residential POI & -0.0639\sym{***} & -0.0474\sym{***} & -0.0468 & -0.0323\sym{***} & -0.0212\sym{***} \\ 
   & (0.0052) & (0.0050) & (0.0050) & (0.0020) & (0.0021) \\ 
  Taxi & 0.1757\sym{***} & 0.2060\sym{***} & 0.2549 & 0.0480\sym{***} & 0.0250\sym{***} \\ 
   & (0.0043) & (0.0037) & (0.0037) & (0.0007) & (0.0011) \\  \bottomrule
			\multicolumn{5}{l}{\footnotesize{\textsuperscript{1} Coefficients are on the $\log$ scale.}} \\
			\multicolumn{5}{l}{\footnotesize{Standard errors in parentheses. $^{*} p < .05,\, ^{**} p < .01,\, ^{***} p < .001$}}
		\end{tabular}
		\caption{Estimates and standard errors for property crime in the full setting (setting 8).}
		\label{tbla:fullproperty}
	\end{table}

	\begin{table}[h!]
		\small
		\centering
		\sisetup{input-symbols=(), table-number-alignment = center, group-digits=false, table-space-text-post = \sym{***}
		} 
		\begin{tabular}{l*{5}{S[table-format=-1.4, table-align-text-post=false]}}
			\toprule
			Variable& {CAR} & {SAR} & {LR} & {GLM\textsuperscript{1}} & {GLMM\textsuperscript{1}}  \\ \midrule
Intercept & -0.6394\sym{***} & -0.8085\sym{***} & -0.8674\sym{***} & -3.2569\sym{***} & -3.4166\sym{***} \\ 
   & (0.0863) & (0.0780) & (0.0785) & (0.1781) & (0.0001) \\ 
  Population & 0.0000\sym{**} & 0.0000\sym{***} & 0.0001\sym{***} & 0.0001\sym{***} & 0.0001\sym{***} \\ 
   & (0.0000) & (0.0000) & (0.0000) & (0.0000) & (0.0000) \\ 
  Median age & -0.0009 & -0.0013\sym{*} & -0.0019\sym{**} & -0.0258\sym{***} & -0.0063\sym{***} \\ 
   & (0.0008) & (0.0006) & (0.0006) & (0.0020) & (0.0004) \\ 
  Male & 0.7738\sym{***} & 1.0597\sym{***} & 1.1766\sym{***} & 2.0673\sym{***} & 1.2500\sym{***} \\ 
   & (0.1386) & (0.1302) & (0.1311) & (0.2818) & (0.0001) \\ 
  Black & 0.2058\sym{***} & 0.0412 & 0.0991\sym{***} & 1.3403\sym{***} & 1.5275\sym{***} \\ 
   & (0.0325) & (0.0227) & (0.0228) & (0.0594) & (0.0003) \\ 
  Asian & 0.0773\sym{*} & -0.0142 & -0.0094 & 0.6724\sym{***} & 1.1663\sym{***} \\ 
   & (0.0376) & (0.0242) & (0.0244) & (0.0760) & (0.0000) \\ 
  Hispanic & 0.3434\sym{***} & 0.1154\sym{***} & 0.1931\sym{***} & 1.2196\sym{***} & 1.7608\sym{***} \\ 
   & (0.0363) & (0.0264) & (0.0266) & (0.0646) & (0.0003) \\ 
  Vacancy rate & 0.3272\sym{***} & 0.4673\sym{***} & 0.5500\sym{***} & 1.3155\sym{***} & 0.2863\sym{***} \\ 
   & (0.0735) & (0.0617) & (0.0621) & (0.1385) & (0.0001) \\ 
  Female-headed HH & 0.8420\sym{***} & 1.4462\sym{***} & 1.6264\sym{***} & 1.8172\sym{***} & 0.3705\sym{***} \\ 
   & (0.0853) & (0.0721) & (0.0726) & (0.1605) & (0.0002) \\ 
  log night tweets & 0.0107\sym{**} & 0.0104\sym{***} & 0.0158\sym{***} & 0.1099\sym{***} & 0.0207\sym{***} \\ 
   & (0.0035) & (0.0029) & (0.0030) & (0.0065) & (0.0053) \\ 
  Entertainment POI & -0.0005 & 0.0008 & 0.0027\sym{*} & -0.0056 & -0.0072\sym{**} \\ 
   & (0.0013) & (0.0013) & (0.0013) & (0.0031) & (0.0024) \\ 
  Uni POI & 0.0007 & 0.0021 & 0.0027\sym{*} & 0.0086\sym{**} & 0.0055\sym{*} \\ 
   & (0.0011) & (0.0011) & (0.0011) & (0.0028) & (0.0026) \\ 
  Food POI & 0.0059\sym{***} & 0.0073\sym{***} & 0.0070\sym{***} & 0.0154\sym{***} & 0.0245\sym{***} \\ 
   & (0.0016) & (0.0016) & (0.0016) & (0.0036) & (0.0029) \\ 
  Professional POI & 0.0068\sym{***} & 0.0046\sym{*} & 0.0045\sym{*} & 0.0189\sym{***} & 0.0181\sym{***} \\ 
   & (0.0019) & (0.0019) & (0.0019) & (0.0045) & (0.0035) \\ 
  Nightlife POI & 0.0035\sym{*} & 0.0026 & 0.0030 & 0.0149\sym{***} & 0.0081\sym{**} \\ 
   & (0.0017) & (0.0017) & (0.0017) & (0.0037) & (0.0030) \\ 
  Outdoors POI & 0.0016 & -0.0028 & -0.0028 & -0.0053 & 0.0001 \\ 
   & (0.0021) & (0.0020) & (0.0020) & (0.0047) & (0.0040) \\ 
  Shops POI & 0.0041\sym{*} & 0.0006 & 0.0001 & -0.0056 & 0.0129\sym{***} \\ 
   & (0.0017) & (0.0017) & (0.0017) & (0.0039) & (0.0033) \\ 
  Travel POI & 0.0034 & 0.0014 & -0.0004 & -0.0118\sym{**} & 0.0280\sym{***} \\ 
   & (0.0018) & (0.0017) & (0.0017) & (0.0039) & (0.0031) \\ 
  Residential POI & -0.0058\sym{**} & -0.0029 & -0.0027 & -0.0026 & -0.0091\sym{**} \\ 
   & (0.0018) & (0.0017) & (0.0018) & (0.0042) & (0.0033) \\ 
  Taxi & 0.0530\sym{***} & 0.0877\sym{***} & 0.1094\sym{***} & 0.1205\sym{***} & 0.0754\sym{***} \\ 
   & (0.0054) & (0.0049) & (0.0049) & (0.0060) & (0.0073) \\ 
		\bottomrule
		\multicolumn{5}{l}{\footnotesize{\textsuperscript{1} Coefficients are on the $\log$ scale.}} \\
		\multicolumn{5}{l}{\footnotesize{Standard errors in parentheses. $^{*} p < .05,\, ^{**} p < .01,\, ^{***} p < .001$}}
	\end{tabular}
		\caption{Estimates and standard errors for violent crime in the full setting (setting 8)}
		\label{tbla:fullviolent}
	\end{table} 
	Since the significance levels vary across models, we do not discuss each model individually. Instead, we focus on effects identified as significant by all models and refer to the average effect over models in the text. As the coefficients for GLM and GLMM are on the log-scale, we present the effects for linear and exponential models separately. 
	
	For property crime, the largest effect size across all non-exponential models is observed for the vacancy rate, which is significantly positively associated with property crime counts. 	
	The new features are significantly associated with property crime. In particular, a 1 unit increase in the weekly taxi flow is associated with an increase of 0.21 property crime counts. Similarly, an increment of one venue in the shops category results in a 0.13 increase of crime counts. Interestingly, a single additional residential venue, often elderly homes, is associated with a 0.05 decrease of property crime. This is an intuitive result when considering the higher presence of watchful neighbours. A similar result is observed for nightlife venues, which are associated with a 0.07 decrease. While the Twitter feature is significant, its effect of property crime is comparatively small as a 1 percent increase in night tweets yields a $0.14/100 = 0.0014$ increase in crime counts. 
        For the exponential models, we observe very similar results. The largest effect is, again, observed for the vacancy rate, followed by the taxi feature. The same POI venues are identified as influencing property crime counts.
	
	For violent crime, the effect of social cohesion is pronounced. A 10\% increase in the male share predicts a 0.1 increase of violent crime counts. Similarly, 10\% increases in the rates of female-headed households and vacant homes are associated with increases of 0.1 and 0.04 in counts. The relevance of ethnic heterogeneity is less pronounced compared to property crime. 
    
     \begin{table}[h!]
    \centering
    \resizebox{.6\textwidth}{!}{%
      \begin{tabular}{cllll}
       \toprule
        \makecell[l]{Rank} & RF & \makecell[l]{Mean\\rank} & GBM &\makecell[l]{Mean\\rank}\\ \midrule
 1. & Taxi & 1.00 & Taxi & 1.00 \\ 
 2. & log night tweets & 2.00 & Hispanic & 2.00 \\ 
 3. & Hispanic & 3.00 & Entertainment POI & 3.00 \\ 
 4.& Population & 4.25 & Median age & 3.50 \\ 
 5. & Shops POI & 4.50 & log night tweets & 4.08 \\ \bottomrule
      \end{tabular}%
}
    \caption{Variable importance for property crime in Setting 8 over 13 windows}\label{tbl:varimp_property}
    \end{table}
	
	Regarding the new features, the effect of the Twitter feature is even smaller than for property crime. This is contrasted with the taxi feature where a 1 unit increase yields a 0.08 increase in violent crime. As with property crime, the food category has the largest effect size. Even then, an additional food venue is associated with a relatively small increase of 0.007 in violent crime.
    Again, the results for the exponential models are similar. The largest effects over both models are observed for demographic variables such as the male share of female-headed households. 
    
	With respect to spatial dependence, we find that estimates of the corresponding parameter in the CAR model are considerably larger than in the SAR model. For the CAR model, the average estimate for $\delta$ is $0.1357$. For the SAR model, we obtain an average $\rho$ estimate of 0.0629. Since the CAR model implies stronger local autocorrelation we find evidence for substantial dependence on direct neighbours. 
    
\begin{table}[h!]
\centering
\resizebox{.6\textwidth}{!}{%
\begin{tabular}{cllll} \toprule
\makecell[l]{Rank} & RF & \makecell[l]{Mean\\rank} & GBM & \makecell[l]{Mean\\rank}\\ \midrule 
   1. & Taxi & 1.00 & Taxi & 1.00 \\ 
  2. & log night tweets & 2.00 & Female-headed HH & 2.00 \\ 
  3. & Female-headed HH & 3.08 & Population & 3.08 \\ 
  4. & Population & 4.25 & log night tweets & 3.92 \\ 
  5. & Black & 4.64 & Median age & 5.00 \\ \bottomrule 
 \end{tabular}%
}
\caption{Variable importance for violent crime in Setting 8 over 13 windows}\label{tbl:varimp_violent}
\end{table}	
		
    We complete the explanatory analysis by inspecting the variable importance for the machine learning models. Since we estimate models over 13 windows, we average the importance rank for each variable over 13 windows. We present the five variables with the overall highest mean ranks in Tables~\ref{tbl:varimp_property} and \ref{tbl:varimp_violent}. If the mean rank equals the importance rank, the variable has that rank across all 13 windows. We find that for both crime types, the taxi feature is highly important. Furthermore, the Twitter feature, which does not have a large effect size in the econometric models, is highly ranked for both crime types and machine learning models. Overall, we find that the regression results and the variable importance ranking are in agreement.
    
\begin{figure}
	\centering
    \includegraphics[width=.55\textwidth]{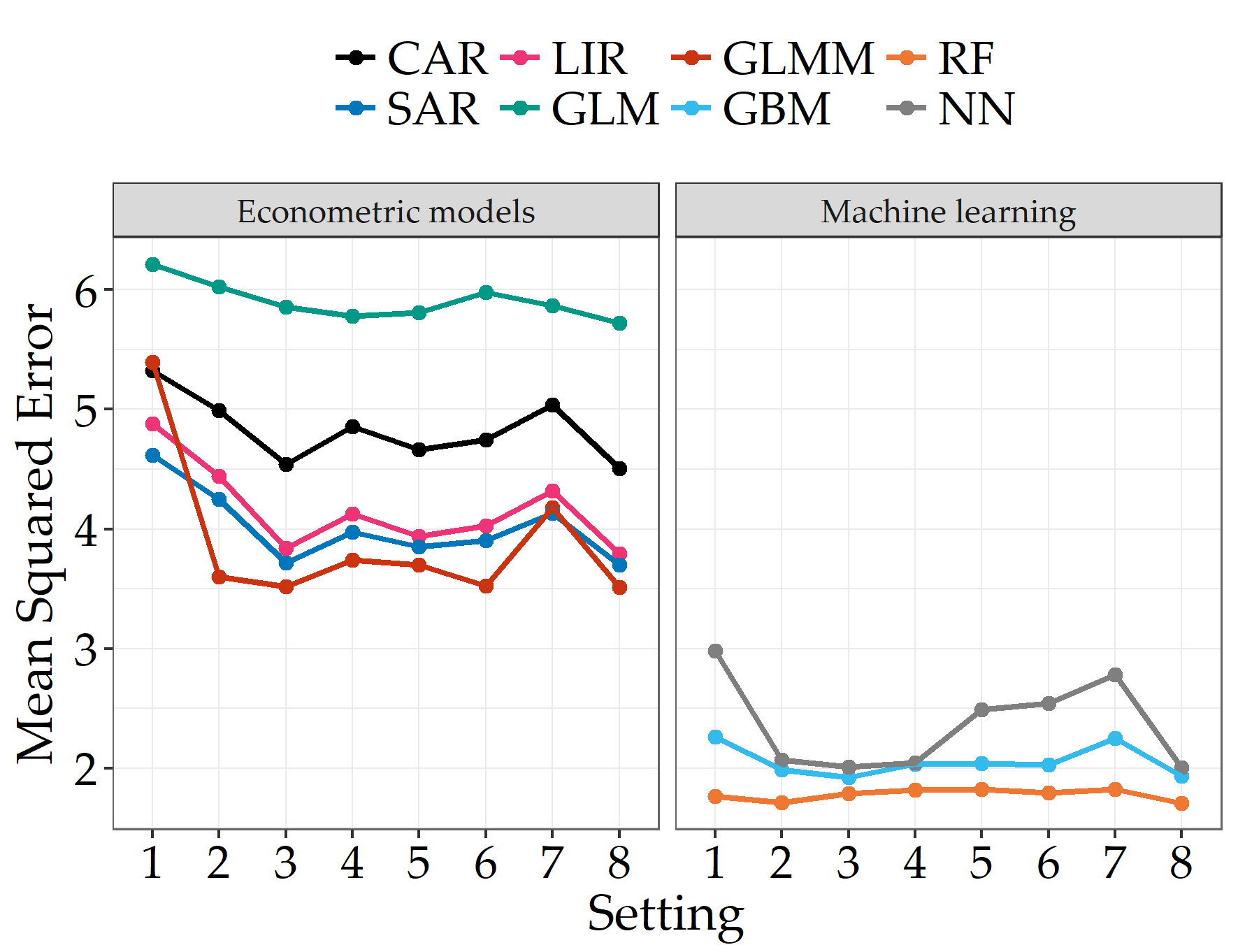}
    \caption{MSE values of different models for property crime predictions.}
    \label{fig5:propertyerrors}
\end{figure}

\begin{figure}
	\centering
    \includegraphics[width=.65\textwidth]{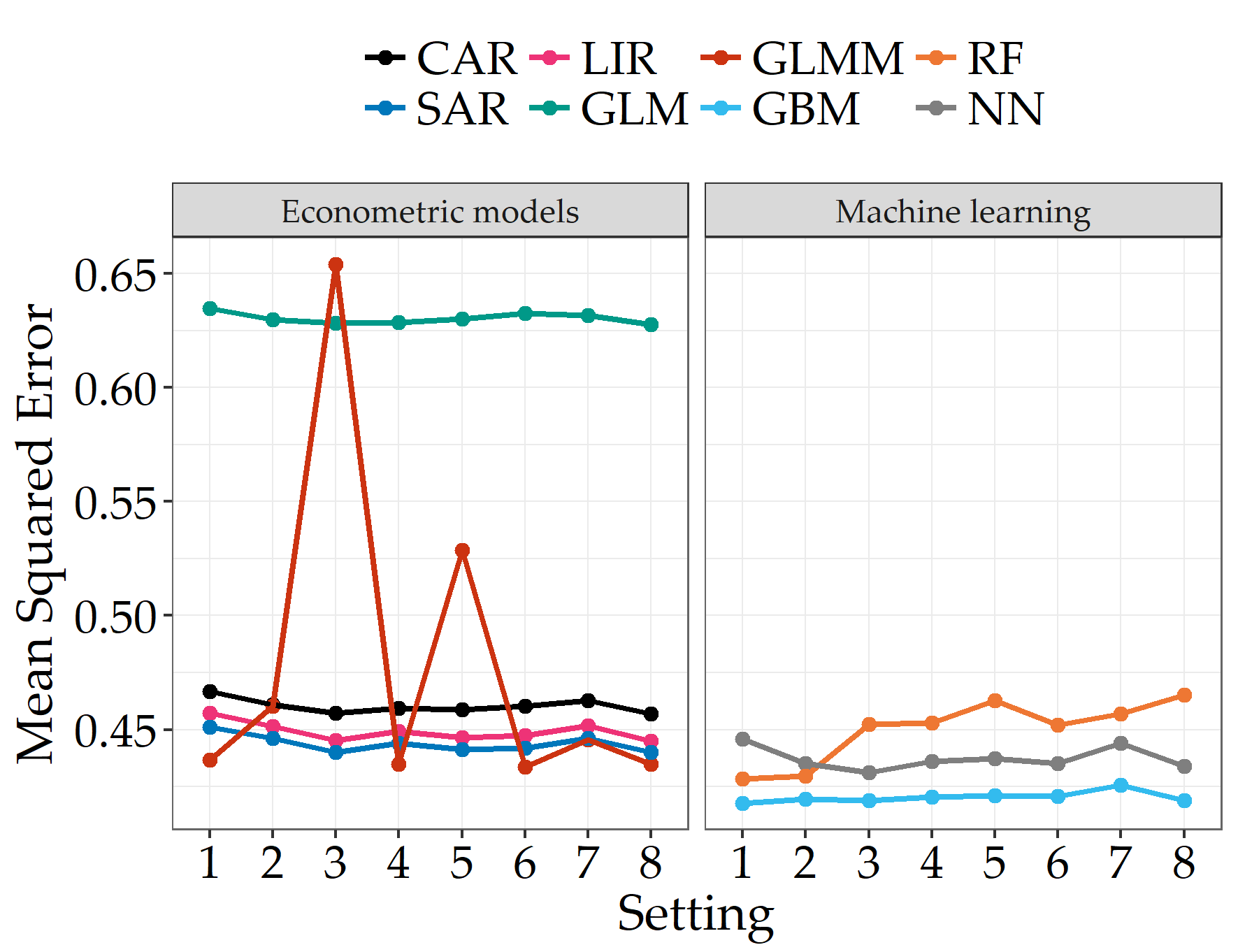}
    \caption{MSE values of different models for violent crime predictions.}
    \label{fig5:violenterrors}
\end{figure}

    We now focus on the predictive results. Figures~\ref{fig5:propertyerrors} and \ref{fig5:violenterrors} plot the MSE over 13 periods. 
    For property crime, we observe a clear pattern: the MSE is largest across all models for setting 1, which uses demographic variables only, and it decreases upon adding novel features. This provides strong evidence in favour of using novel data sources for property crime prediction. In addition, we observe that some features perform better when used in combination. In particular, settings 3 and 5 use the taxi feature together with POI data (setting 3) or Twitter (setting 5). These settings perform better than the combination of POI data and Twitter data alone. Adding only one feature already improves the predictive accuracy but to a lesser degree compared with adding a combination. Setting 8 using all features together produces the best result. Over all models considered, the MSE in setting 8 is on average 19\% lower compared to the baseline setting. This is the largest improvement compared to all other settings, which result in a MSE that is on average 11\% lower than the baseline. 
    Clearly, the machine learning models outperform the econometric models across all settings. A Random Forest in setting 8 produces the smallest prediction error. We suggest that the superior performance is driven by non-linear relationships between the features and property crime.

    For violent crime, there is a very different trend. As before, the machine learning models perform better than the econometric models but the margin is smaller. With respect to novel data sources, the econometric models slightly improve on their predictions in the baseline setting (setting 1) when having access to the full set of features (setting 8). The machine learning techniques, however, benefit very little from the new features. We observe the lowest prediction error with a GBM in Setting 1. Over all models, the MSE in the other settings is 1\% higher than the MSE in setting 1. We conclude that using data on human dynamics and POI offers little advantage for violent crime prediction.

\begin{figure}
\centering
\begin{subfigure}{.45\textwidth}
\centering
\includegraphics[width=.9\linewidth]{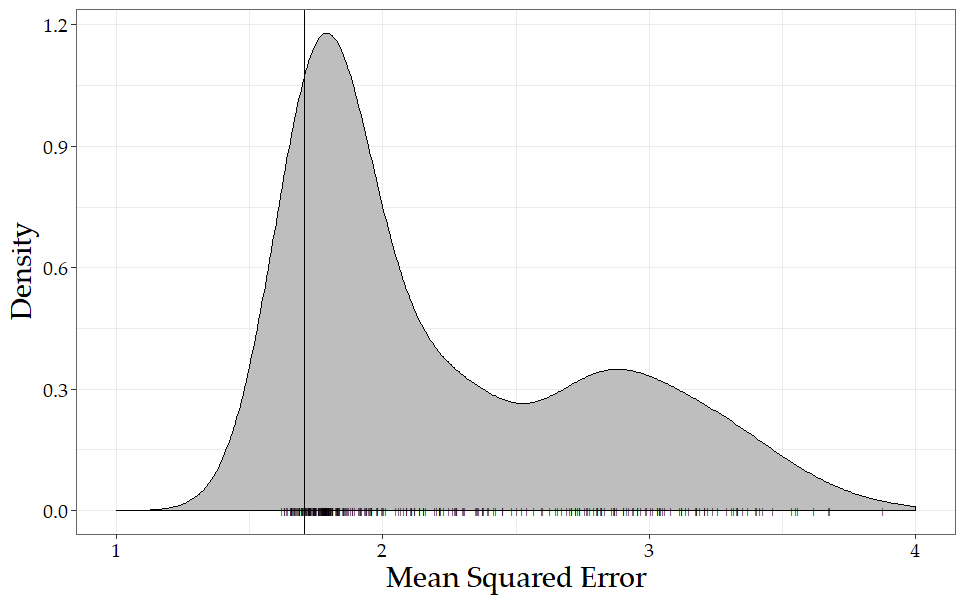}
\caption{Property crime: RF in setting 8.}
\end{subfigure}%
\begin{subfigure}{.45\textwidth}
\centering
\includegraphics[width=.9\linewidth]{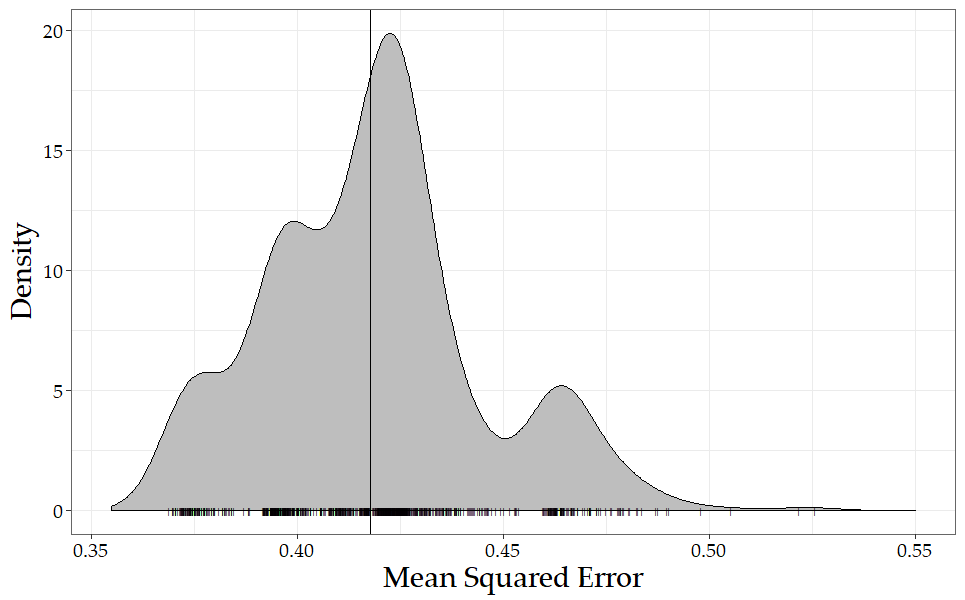}
\caption{Violent crime: GBM in setting 1}
\end{subfigure}
\caption{MSE distribution over hyperparameters and windows}\label{fig5:dens}
\end{figure}

	We investigate the robustness of our results for the two best performing models: a RF using setting 8 for property crime and a GBM for violent crime in setting 1. In Figure \ref{fig5:dens}, we plot the MSE obtained for individual windows hyperparameter configurations during grid search. Each point on the x-axis corresponds to a MSE obtained for a single window and hyperparameter configuration. 
    The vertical line corresponds to the lowest average MSE obtained over all windows as reported in Figures \ref{fig5:propertyerrors} and \ref{fig5:violenterrors}. Especially for property crime, there is a clear peak regarding the mean MSE for each individual window which means that the predictions are relatively robust against specific hyperparameter settings as they all yield similar results. This provides strong evidence for the superiority of the new features since a wide range of RF produce competitive property crime predictions. 
    
    The results for the GBM predicting violent crime are different: the prediction errors are more variable as a function of hyperparameters and windows and the best-performing hyperparameter combinations at each window are  more dissimilar than for property crime. Given that these results are obtained with Census data only, the sensitivity to the window choice is not surprising. 

	\section{Discussion}
	\label{sec:discussion}
	Our mixed approach of explanatory analysis and prediction reflects the dual objective of police and policy makers. We can not only show that crime forecasting benefits from including the novel feature, we also shed light on the emergence of urban crime and find clear support for well-known crime theories. This provides clear guidance on how to conceptualise and address crime in a predictive policing context. 
         
    The forecasting results show that using the new features significantly improves the prediction accuracy for property crime. We find that adding static data such as POI venues does not suffice to forecast crime counts accurately. Instead, dynamic Twitter or taxi data and in particular their interaction greatly reduce the prediction error. These results are in line with prior work by \citet{wang2016crime} and \citet{bendler2014crime}.
    We suggest that a combination of node-specific data on the demographic make up as well as the visitor make up through Twitter and POI data in combination with edge-specific data on social taxi flow is the best combination of different data sources to predict property crime counts. The taxi feature proxies human dynamics between areas and how people proliferate crime through space. The spatial dependence matrix models only first-order dependence of immediate neighbours. Many taxi trips traverse multiple areas such that the taxi feature accounts for social connection and crime proliferation beyond just neighbouring sites.
	
	For violent crime, however, the spatio-temporal dimension of the new features adds very little. 
    Our explanatory analysis reveals the origins of this result. Violent crime is taking place in neighbourhoods with poor social cohesion as evident by the positive association with vacant homes and female-headed family households. In line with disorganisation theory, social deprivation provides the context for delinquent, violent behaviour \citet{kubrin2003new}. Support for social disorganisation theory is supplemented by the fact that violent crime counts are not particularly sensitive to POI venues. 
    That long-term structural conditions are more important for violent crime is further emphasised by the poor explanatory and predictive performance of short-term human activity as captured by the novel features.
    
    In contrast, property crime is far less related to the residential make up of the census tract where the crime takes place. Rather than local deprivation, local opportunities through anonymity and vacant homes matter. The coefficients and variable importance rankings capture a trade-off between more opportunities and targets through high human activity on the one hand and more watchful eyes, deterring crime on the other hand. This is for instance illustrated in the negative association of property crime with nightlife and residential venues and the positive association with shopping venues. 
    The notion that different circumstances drive property and violent crime differently is further supported by the rather low correlation between the crime types (Pearson's $r = 0.17$), indicating that the two crime types take place in areas with very different characteristics.
    
    Police react to crime with temporary resource allocations as well as with long-term policy decisions on funding, intervention programmes and task forces. In order to tackle crime, public decision makers depend on both immediate, accurate crime volume forecasts and insight into the underlying crime generating process.    
    
  Our explanatory analysis reveals that violent crime emerges from long-standing social environments where short-run movement dynamics do not matter. 
	Based on these results, crime prevention strategies need to account for this spatial and structural difference. Since violent crime is a more slowly-varying process, corresponding crime prevention programmes need to address long-standing issues of re-victimisation and re-offending through youth and family support programmes and partnerships with affected communities. 
	
	In contrast, our results indicate that to prevent property crime, police need to be aware of its transitory, changing nature. It is driven by localised opportunities, which means that interventions need to target those intersections of opportunity and offender. In particular, the relevance of the taxi feature demonstrates that in large cities, both offenders and victims cross large distances, propagating crime. This implies that police need to consider not only neighbouring areas but also connections to areas that are further away.
	Anonymous data on human behaviour can be crucial in identifying these links. 
    
    At the time of study, the limited availability of Twitter data constrained the time period of study. Future research can exploit different sources of social media activity and investigate whether similar results hold outside of the United States.
    	
	\section{Conclusion}
	\label{sec:conclusion}
	This paper presents a multi-model solution to predicting the number of crime incidents in a census tract by combining demographic data with aggregated social media, venue and taxi flow data. 
	In addition, it addresses the two-fold concerns of policy makers: preventing crime in the short run through resource allocation and preventing crime in the medium run through prevention programmes.
    
      Using a rolling-window prediction approach, we provide robust evidence that new features accounting for human activity improves forecasts for crimes shaped by local opportunities. 
	By not only relying on previous crime observations and quinquennial census data but rather on abundantly available behavioural data, the models can generalise to new areas or areas with poor reporting rates. 
		
        Following an applied perspective, the proposed approach can be employed to predict future problematic crime areas and improve police responsiveness and resource allocation. By analysing underlying mechanisms of different crime types, promising areas for intervention have been identified. 
	
   \section*{Funding}
   \noindent
   This research did not receive any specific grant from funding agencies in the public, commercial, or not-for-profit sectors.

	\bibliography{main}

	\appendix
	    
	\section{Grid Search Parameters}\label{ap:grid}
    Table~\ref{tbl:hyper} details which parameters were optimised during a grid search. We use early stopping when the MSE does not decrease by at least 0.01\% for 5 consecutive scores. Where different, we supply the values used for property and violent crime fitting separately.
    
    \begin{table}
    \centering
\begin{tabular}{lp{5cm}p{5cm}l} \toprule
Model & Parameter & \multicolumn{2}{c}{Range of values} \\ \cmidrule{3-4}
& & Property & Violent \\
  \midrule
\multirow{10}{*}{GBM} & Learn rate & {0.01--0.2 with 0.01 increments} \\ 
  & Learn rate annealing & {0.990--0.998 with 0.001 increments} \\ 
  & Maximum allowed tree depth & 13--21 & 7--15\\ 
  & Row sample rate & {0.20--1 with 0.05 increments} \\ 
  & Column sample rate & {0.20--1 with 0.05 increments} \\ 
  & Column sample rate per tree & {0.20--1 with 0.05 increments} \\ 
  & Minimum number of rows in a terminal node & {4, 8, 16, 32, 64, 128, 256, 512} \\ 
  & Number of bins used for split & {16, 32, 64, 128, 256, 512, 1024} \\ 
  & Error improvement threshold for split & {0, $10^{-8}, 10^{-6}, 10^{-4}$} \\ 
  & Histogram type at each node & {Quantiles Global, Round Robin} \\ 
  & Number of trees &{10,000} \\ \midrule
  \multirow{5}{*}{NN} & Learning rate & {adaptive (ADADELTA)} \\
  & Neurons in hidden layer(s) & {64, 128, 256, 512} \\ 
  & Number of hidden layers & {1, 2} \\
  & Epochs & {1, 10, 20} \\ 
  & Learning rate decay & {0.95, 1 (no decay)} \\ 
  \midrule
  \multirow{8}{*}{RF} & Maximum allowed tree depth & 11--19 & 7--15 \\ 
   & Row sample rate & {0.20--1 with 0.05 increments}\\ 
   & Column sample rate & {0.20--1 with 0.05 increments}\\ 
   & Minimum number of rows in a terminal node & {4, 8, 16, 32, 64, 128, 256, 512} \\ 
   & Number of bins used for split &{16, 32, 64, 128, 256, 512, 1024}\\ 
   & Error improvement threshold for split & {0, $10^{-8}, 10^{-6}, 10^{-4}$} \\ 
   & Histogram type at each node & {Quantiles Global, Round Robin}\\ 
  & Number of trees & {10,000}\\ \bottomrule
\end{tabular}
\caption{Range of grid search values for hyperparameter optimisation}\label{tbl:hyper}
\end{table}

    \section{Coefficients in rolling window estimation} \label{ap:coef}
    Since we re-estimate the linear models in each window, we obtain a distribution of coefficients over 13 windows. Since setting 8 includes all variables, we present the coefficients for all models for setting 8.
    
    \begin{figure}
    \centering
    \includegraphics[width = .8\textwidth]{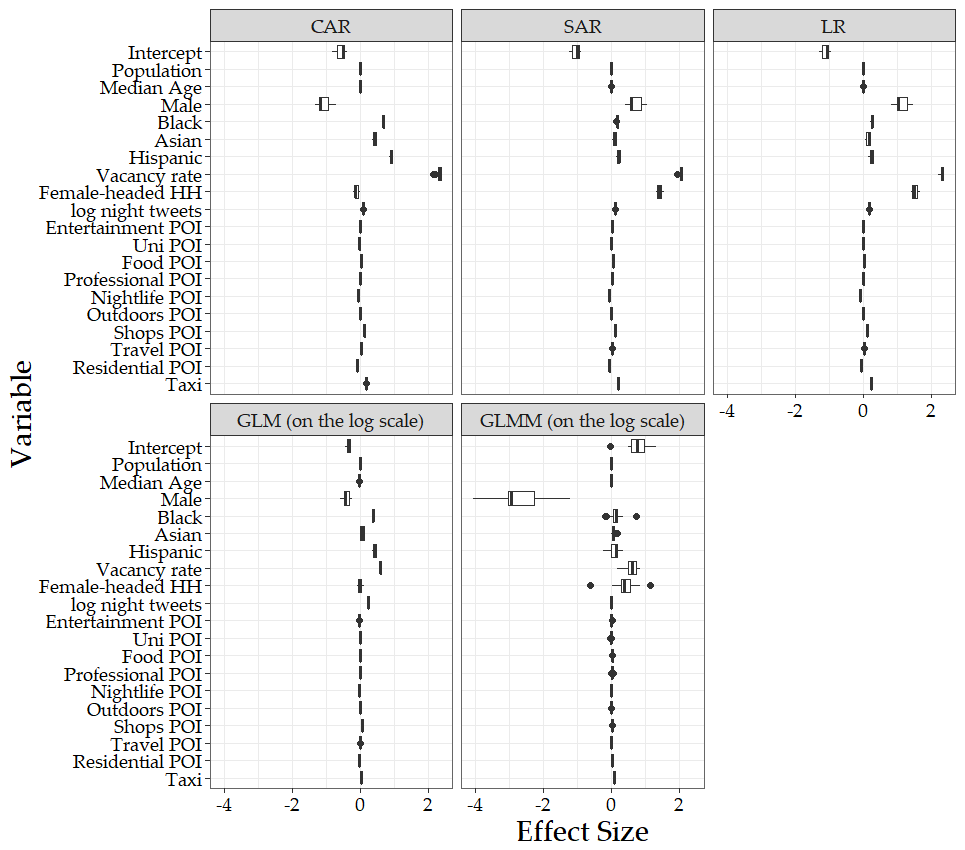}
    \caption{Coefficient distribution for property crime for Setting 8 over 13 windows.}\label{fig:apcoef_property}
    \end{figure}
    
        \begin{figure}
    \centering
    \includegraphics[width = .8\textwidth]{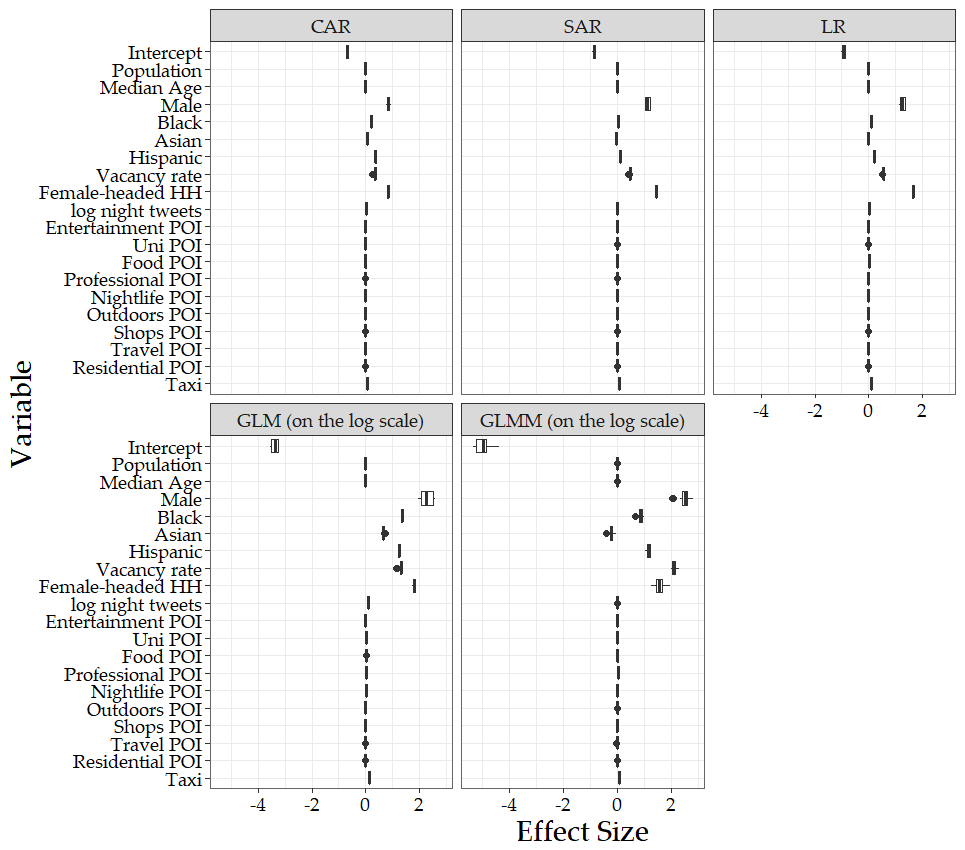}
    \caption{Coefficient distribution for violent crime for Setting 8 over 13 windows.}\label{fig:apcoef_violent}
    \end{figure}
	
\end{document}